\documentclass[conference]{IEEEtran}
\ifCLASSINFOpdf
\else
\fi
\usepackage[rightcaption]{sidecap}
\usepackage{graphicx} 
\usepackage{multirow}
\usepackage{arydshln}

\usepackage{floatrow}
\newfloatcommand{capbtabbox}{table}[\captop][\FBwidth]

\newcommand{\RomanNumeralCaps}[1]
    {\MakeUppercase{\romannumeral #1}}

\usepackage{float}
\floatstyle{plaintop}
\restylefloat{table}

\IEEEoverridecommandlockouts
\usepackage{cite}
\usepackage{amsmath,amssymb,amsfonts}
\usepackage{algorithmic}
\usepackage{graphicx}
\usepackage{float}
\usepackage{subfigure}
\usepackage{textcomp}
\usepackage{xcolor}
\def\BibTeX{{\rm B\kern-.05em{\sc i\kern-.025em b}\kern-.08em
    T\kern-.1667em\lower.7ex\hbox{E}\kern-.125emX}}
\begin{document}
%
\title{Zebra: Memory Bandwidth Reduction for CNN Accelerators With Zero Block Regularization of Activation Maps 
}

\author{\IEEEauthorblockN{Hsu-Tung Shih and Tian-Sheuan Chang}
\IEEEauthorblockA{\textit{Dept. of Electronics Engineering, 
 National Chiao Tung University
Hsinchu, Taiwan}}
\thanks{
H. -T. Shih and T. -S. Chang, "Zebra: Memory Bandwidth Reduction for CNN Accelerators with Zero Block Regularization of Activation Maps," 2020 IEEE International Symposium on Circuits and Systems (ISCAS), 2020, pp. 1-5, doi: 10.1109/ISCAS45731.2020.9180865.
}
}

\maketitle

\begin{abstract}
 The large amount of memory bandwidth between local buffer and external DRAM has become the speedup bottleneck of CNN hardware accelerators, especially for activation maps. To reduce memory bandwidth, we propose to learn pruning unimportant blocks dynamically with zero block regularization of activation maps (Zebra). This strategy has low computational overhead and could easily integrate with other pruning methods for better performance. The experimental results show that the proposed method can reduce 70\% of memory bandwidth for Resnet-18 on Tiny-Imagenet within 1\% accuracy drops and 2\% accuracy gain with the combination of Network Slimming.

\end{abstract}


\IEEEpeerreviewmaketitle

\section{Introduction}
Convolutional neural networks (CNNs) have been the state of the art approach in many computer vision tasks, including image recognition \cite{simonyan2014very, he2016deep}, object detection, etc. However, CNNs are often over-parameterized, causing waste of computation times and energy. Therefore, a number of model compression techniques have been proposed to solve this problem. The proposed method can be separated into two main groups, static pruning and dynamic pruning.

Static pruning prunes the model weights without taking input data into consideration. Weight pruning 
\cite{han2015learning} prunes the most unimportant weight based on magnitude to reduce model size, but it needs specific CNNs accelerator to reduce execution time. Network Slimming \cite{liu2017learning} prunes the whole channel based on the scaling factor gamma in a batch normalization (BN) \cite{ioffe2015batch} layer to get a more structured model. 

Dynamic pruning, on the other hand, prunes different subsets of model with different input data. \cite{gao2018dynamic,channelgating} dynamically selects important channels in run-time and won't execute those  unimportant ones. However, both of them require additional branch structures and complicate the training process. 


\begin{figure}
\begin{floatrow}

\ffigbox{%

\includegraphics[width=0.45\textwidth]{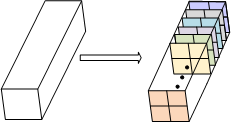}
}{%
  \hfill
  
  \caption{Illustration of separating activation maps into several non-overlapping blocks. The number of blocks depends on the block size and the activation maps size. Original model has block size = whole map.}%
  \label{fig:blocks}
}

\capbtabbox{%
\caption{Percentage of Zero blocks of Resnet-18 on CIFAR-10.}%
  \begin{tabular}{cc} \hline
  
  Block size & Results \\ \hline
  $2\times 2$ & $24.7\%$ \\
  $4\times 4$ & $7.9\%$ \\ 
  whole map& $1.1\%$ \\ \hline
  \end{tabular}
  \label{tab:blocks}
}{%
  
}
\end{floatrow}
\end{figure}

Most of the existed static and dynamic pruning methods focus on reduction of  FLOPS or model size instead of  memory bandwidth. However, memory bandwidth has gradually become the bottleneck in modern convolution neural networks system since data transfer between local and global memory wastes lots of time and power. These data include activation maps and model weights. Usually, the size of activation maps is several order of magnitudes larger than the model size \cite{lin2017data,chen2016eyeriss}. In other words, reducing the size of activation maps will be effective to solve the memory bandwidth problem. 

A common approach to reduce the size of activation maps uses ReLUs \cite{nair2010rectified} to force negative part to be zero to generate sparse activation maps. However, the zero distribution of such approach is irregular, which is bad for compressing maps or skipping computation.  It is hard to simply prune parts of activation maps in structured ways to get real memory bandwidth saving. Besides, the unimportant input such as background in input images will somehow propagate to the activation maps. In \cite{liang2018dynamic}, the authors analyze the sparsity of activation maps in different layers. The result shows that the deeper activation maps have higher zero ratio. In their work, they check activation maps in run-time and prune those with all zero entries. However, this method only reduces little amount of activation maps size due to very few all zero activation maps for large size maps. 

With above observations, we propose a dynamic activation map reduction method to reduce memory bandwidth. Inspired by \cite{liang2018dynamic}, we prune blocks of zeros in the maps as shown in Fig.~\ref{fig:blocks} instead of whole maps in \cite{liang2018dynamic} or fine grained case as introduced by ReLUs. If all values in one block are all zero, we consider it as background or unimportant one and thus we can prune it in run-time. However, the zero block percentage after ReLUs is still quite low as shown in Table.~\ref{tab:blocks}. To force the network learns more "background" or "zero blocks", we propose a regularization method called \underline{Z}\underline{e}ro \underline{B}lock \underline{r}egularization of \underline{a}ctivation maps (Zebra) to achieve dynamic activation map reduction. Our method divides the activation maps into several non-overlapping blocks, then lets the model to dynamically learn unimportant blocks through Zebra, and simply forces all values in those blocks to zero to achieve simple run-time structured activation maps pruning.


The rest of this paper is organized as follows. The proposed regularization techniques Zebra is
detailed in Section \RomanNumeralCaps{2}. Then, experimental results is shown in Section \RomanNumeralCaps{3}. Section \RomanNumeralCaps{4
} concludes this work 


\begin{figure}[t!]

\includegraphics[width=1.0\textwidth]{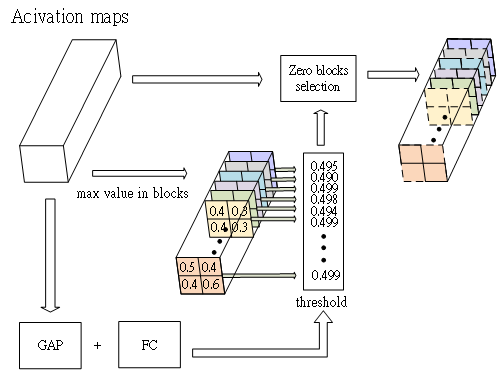}
    \caption{Zebra in the training mode, where GAP is a global average pooling layer. $T_{obj}$ is set to 0.5 in this example.}
	\label{fig:training}
\end{figure}

\section{Proposed method}

\subsection{Model Training}
Fig.~\ref{fig:training} shows the flow of Zebra to train zero blocks. Activation maps are first divided into non-overlapped blocks. Zebra learns zero blocks or unimportant blocks through regularization. The importance of a block is represented by the maximum value in each block. If the maximum value of a block is smaller than a learned threshold, this block will be regarded as a zero block. The blocks in the same channel use the same threshold. The threshold is learned by feeding activation maps to a small network with a global average pooling layer and a fully-connected layer. 

 Based on above flow, the sparsity of blocks is controlled by introducing a regularization term into the overall loss function as  
\begin{equation}
L=\lambda Loss_{CE} + \sum_{l,c}\| T_{obj}-T_{l,c}\|^2
\end{equation}
In which, the first term is the cross entropy loss used in image classification problem. The second term is our Zebra regularization term, where $l$ and $c$ represents different layers and channels respectively and $\lambda$ is used to balance these two terms. We use $L_2$ loss to force the threshold of each layer and channel $T_{l,c}$ close to the target threshold $T_{obj}$. $T_{obj}$ value can be typically set to  a value within [0,1] to get different level of blocks sparsity since  the values in activation maps are almost in the range of [0,1] with the help of the BN layer. Empirically, $T_{obj}$ is decided based on simple trials.

\begin{figure}[t!]

\includegraphics[width=1.0\textwidth]{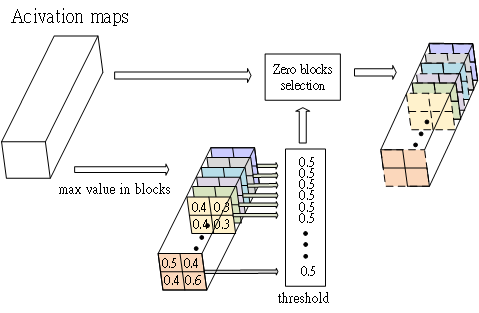}
    \caption{Zebra in the inference mode. $T_{obj}$ is set to 0.5 in this example and $T_{l,c}$ will be converged to $T_{obj}$ at the end of training. Therefore, we can directly set $T_{l,c}$ to $T_{obj}(0.5)$ for inference.}
	\label{fig:inference}
\end{figure}

\subsection{Model Inference}
After training, we examine the threshold values obtained by Zebra regularization. To our surprise, the learned threshold values are almost converged to the given $T_{obj}$ as shown in Fig.~\ref{fig:inference}. Therefore, during inference we can remove the small network and just use the user-defined $T_{obj}$ as the threshold in run-time as shown in Fig.~\ref{fig:inference}.
    
\subsection{Design Parameters Analysis of Zebra}

In Zebra, one important parameter is the block size, which depends on the size of activation maps and datasets. Theoretically, small block size will have higher accuracy and higher sparsity since small block means more fine-grained partitioned. However, small block size will need more index parameters to store the zero conditions of the blocks which leads to memory bandwidth overhead. 

The storage size for memory bandwidth overhead and \textsl{S\%} reduced activation maps with shape $\textsl{C}\times\textsl{W}\times\textsl{H} $ and $\textsl{B}$-bits per element are
\begin{equation}
Activation\ maps\ storage= {C\times W\times H\times B\times S\%  }
\end{equation}
\begin{equation}
Memory\ bandwidth\ overhead= {C\times W\times H \over {block\_size}^2}
\end{equation}
As indicated in this calculation, once the block size is too small, the index storage overhead will be no longer negligible. Therefore, the block size should be chosen carefully.

\begin{table}[]
\begin{tabular}{ll|ll}
\hline
Model        & $T_{obj}$ & \begin{tabular}[c]{@{}l@{}}Reduced \\ bandwidth (\%)\end{tabular} & Test acc. (\%)   \\ \hline
VGG16        & 0        & 16.7                                                              & 92.58          \\ \hdashline
             & 0.05     & 36.4                                                              & 92.35          \\
+ NS (50\%)  & 0.05     & \textbf{51.4}                                                     & 92.40          \\
+ NS (20\%)  & 0.05     & 41.1                                                              & 92.69          \\
+ WP (20\%)  & 0.05     & 42.3                                                              & \textbf{93.27} \\ \hdashline
             & 0.1      & 45.0                                                              & \textbf{92.15} \\
+ NS (50\%)  & 0.1      & \textbf{73.8}                                                     & 89.20          \\
+ NS (20\%)  & 0.1      & 71.1                                                              & 87.81          \\
+ WP (20\%)  & 0.1      & 73.7                                                              & 90.65          \\ \hdashline
             & 0.15     & 54.3                                                              & 91.72          \\ \hline
Resnet-18     & 0        & 2.8                                                               & 91.33          \\ \hdashline
        & 0.1        & 33.5                                                              & 90.41          \\ \hdashline
             & 0.2      & 40.5                                                              & 89.76          \\
+ NS (20\%)  & 0.2      & 41.4                                                              & \textbf{91.55} \\
+ WP (20\% ) & 0.2      & \textbf{49.2}                                                     & 88.62                    \\ \hline
Resnet-56     & 0        & 7.8                                                               & 92.27          \\
             & 0.05     & 31.8                                                              & \textbf{93.22} \\
             & 0.15     & \textbf{46.4}                                                     & 91.33          \\ \hline
MobileNet    & 0        & 14.4                                                              & \textbf{90.66} \\
             & 0.1      & 35.6                                                              & 90.00          \\
             & 0.15     & 78.8                                                              & 87.92         
\end{tabular}

\caption{Simulation results on CIFAR-10.}
\label{tab:cifar}
\end{table}


Another overhead of Zebra is computation overhead. Since the small network can be removed during inference, the computation overhead is just the maximum operation of each block. Consider an activation map with shape $\textsl{C}\times\textsl{W}\times\textsl{H} $ and a $\textsl{C}\times\textsl{F}\times\textsl{F}\times\textsl{O} $ convolutional kernel with stride $s$, the floating point operations per second (FLOPS) for convolution and Zebra computation overhead are 
\begin{equation}
Convolution\ = {C\times W\times H\times F\times F\times O \over s}
\end{equation}
\begin{equation}
Computation\ overhead\ = {C\times W\times H }
\end{equation}
In which, $\textsl{F}\times\textsl{F}$ is kernel size, and $\textsl{O}$ is the output channel number. The above calculation shows that the computation overhead of Zebra is totally negligible. The required hardware implementation is also small enough to be easily integrated with current accelerators after activation functions.


\section{Experimental Results}
\subsection{Settings}
The proposed method is evaluated on two image classification datasets, CIFAR-10 \cite{krizhevsky2009learning} and Tiny-Imagenet that is the subset of Imagenet \cite{deng2009imagenet}. CIFAR-10 consists of 10 classes of natural images with resolution $32 \times 32$.  Empirically, we choose block size as $4$ by considering both activation maps sparsity and the zero blocks index overhead. Note that we set block size as $2$ when the size of activation maps in deeper layers goes to $2 \times 2 $ in VGG and MobileNets.
Tiny-Imagenet contains 200 classes with resolution $64 \times 64$. We set block size to 8 for better activation maps reduction and the index storage overhead trade-off.

We have tested several well-known convolutional neural networks such as VGG \cite{simonyan2014very}, Resnet \cite{he2016deep} and MobileNets \cite{howard2017mobilenets}. All the networks are trained with standard SGD optimizer with learning rate step decay from 0.1 to 0.001. Also, We follow the standard normalization and weight decay techniques on both datasets for better results. 

We also combine Zebra with unstructured magnitude-based weight pruning (WP) \cite{han2015learning} and structured pruning like Network Slimming (NP) \cite{liu2017learning}. Weight pruning removes the unimportant weights of the model. We simply do weight pruning on a well-trained model and use the remaining weights to train with our method. As for Network Slimming, we follow the sparsity training in \cite{liu2017learning} to regulate $\gamma$ in BN layer first, slim the network with given ratio and then retrain with our method.

\subsection{Results}
Table.~\ref{tab:cifar} and ~\ref{tab:tiny} show the memory bandwidth reduction of activation maps and accuracy with Zebra and its combination with other pruning methods. In which, the percentage inside the parentheses means the pruning percentage by these pruning methods first. From the tables, we can find that the amount of bandwidth reduction depends on the models and the datasets. The activation maps reduction can be up to $54\%$ for VGG16, $34\%$ for Resnet-18, $32\%$ for Resnet-56 and $36\%$ for MobileNets with $<1\%$ accuracy drops on CIFAR-10. As for Tiny-Imagenet, it can be up to $70\%$ on Resnet-18 with $<1\%$ accuracy drops.

\begin{figure}[t!]

\includegraphics[width=1\textwidth]{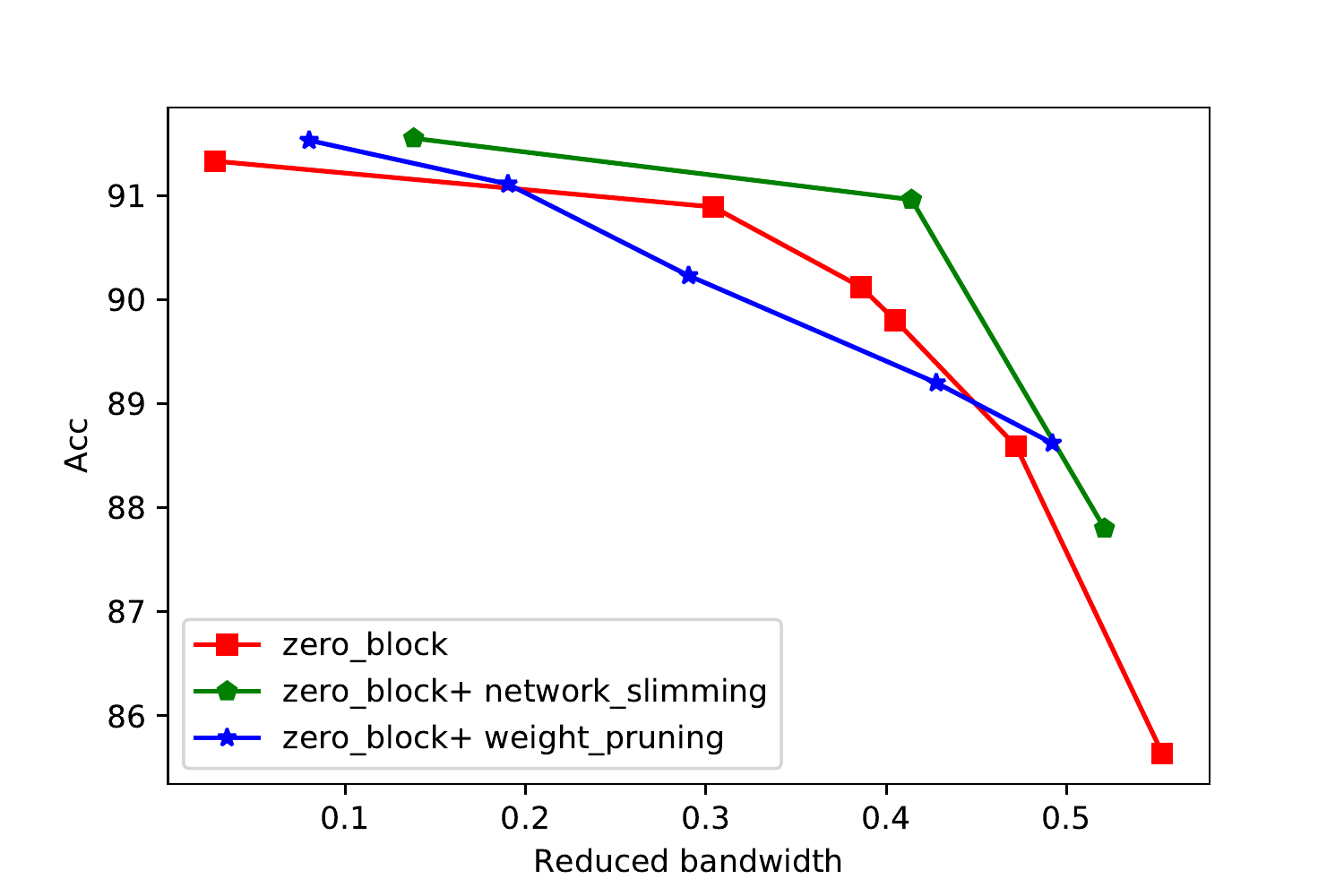}
\caption{Simulation results of Zebra and its combinations with Network Slimming and Weight Pruning for Resnet-18 on CIFAR-10.}
\label{fig:ns}
\end{figure}

When Zebra is combined with Weight pruning, it seems there is no or little performance gain. However, the performance seems to get better when Zebra is used together with Network Slimming. A more detailed ablation study shown in Fig. \ref{fig:ns} and Table. \ref{tab:compare} displays that Network Slimming truly helps Zebra train better. With the help of Network Slimming, the number of redundant activation maps is reduced at the beginning, which makes Zebra training easier.

As to the memory bandwidth overhead of Zebra, the results are shown in Table. \ref{tab:overhead}. The required bandwidth means the total activation maps size for one image and the  bandwidth overhead is the need of block index storage. In this calculation, we assume a layer-by-layer hardware processing that will store the activation maps to external DRAM for each convoluitonal layer processing. This overhead is totally negligible compared with the amount of reduction in activation maps.

\begin{table}[t]
\begin{tabular}{ll|ll}
\hline
Model              & Sparsity & \begin{tabular}[c]{@{}l@{}}Reduced \\ bandwidth (\%)\end{tabular} & \begin{tabular}[c]{@{}l@{}}Test acc. (\%)\\ (top1 / top5)\end{tabular} \\ \hline
Resnet-18           & 0        & 3.0                                                               & 55.18 / 77.56                                                         \\
                   & 0.1      & 15.9                                                              & \textbf{61.46 / 82.50}                                                     \\
                   & 0.15     & \textbf{33.9 }                                                             & 57.00 / 79.64                                                         \\ \hdashline
                   & 0.2      & 47.2                                                              & 56.50 / 78.92                                                         \\
+ NS  (40\%) & 0.2      & \textbf{69.7}                                                              & 58.36 / 79.36                                                         \\
+ NS  (20\%) & 0.2      & 44.5                                                              & \textbf{60.30 / 82.58}                                                         \\
+ WP (40\%) & 0.2      & 41.8                                                              & 59.64 / 81.24                                                           \\
+ WP (20\%) & 0.2      & 42.8                                                              & 58.66 / 80.78                                                         \\ \hdashline
                   & 0.4      & 69.5                                                              & 54.20 / 76.70                                                         
\end{tabular}

\caption{Simulation results on Tiny-Imagenet.}
\label{tab:tiny}
\end{table}

\begin{figure*}[t!]

\includegraphics[width=\textwidth]{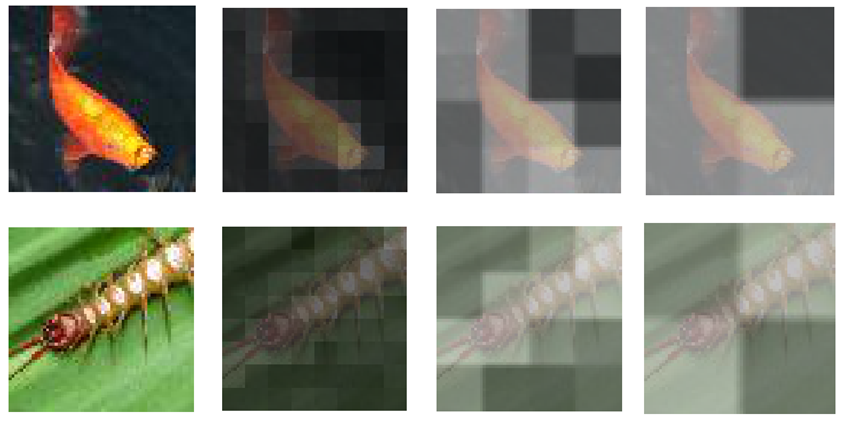}
\caption{Visualization of zero blocks obtained by Zebra for Resnet-18 with Sparsity 0.2 on Tiny-Imagenet. The righter image means the deeper activation maps and we have re-scaled them to the original image size $64 \times 64$. The darker blocks show the more blocks in that situation among channels are zeroed out.}
\label{fig:vis}
\end{figure*}

\begin{table}[t]
\caption{Ablation study of Zebra and its combination with Network Slimming and Weight Pruning on CIFAR-10  }
\begin{tabular}{ll|ll}
\hline
Model     & Method     & \begin{tabular}[c]{@{}l@{}}Reduced \\ bandwidth (\%)\end{tabular} & Test acc. (\%) \\ \hline
VGG16     & NS         & 21.9                                                              & 92.84         \\
          & Zebra      & 40.2                                                              & 92.8          \\
          & Zebra + NS & \textbf{48.5}                                                              & \textbf{92.89}         \\ \hdashline
          & NS         & 58.5                                                              & 90.15         \\
          & Zebra      & 60.4                                                              & 90.23         \\
          & Zebra + NS & \textbf{68.8}                                                              & \textbf{90.25}         \\ \hline
Resnet-18 & NS         & 22.5                                                              & 90.75         \\
          & Zebra      & 30.4                                                              & 90.81         \\
          & Zebra + NS & \textbf{41.4}                                                              & \textbf{90.96}         \\\hdashline
          & NS         & 29.8                                                              & 89.42         \\
          & Zebra      & 40.5                                                              & 89.50         \\
          & Zebra + NS & \textbf{50.4}                                                              & \textbf{89.55}         \\
      
\end{tabular}
\label{tab:compare}
\end{table}

\begin{table}[t]
\begin{tabular}{|l|l|l|l|}
\hline
Model                      & Dataset                                                  & \begin{tabular}[c]{@{}l@{}}Required\\ bandwidth\end{tabular} & \begin{tabular}[c]{@{}l@{}}Bandwidth\\ overhead\end{tabular} \\ \hline
\multirow{2}{*}{Resnet-18} & CIFAR-10                                                 & 2.06 MB                                                      & 4.13 KB (0.2\%)                                                   \\ \cline{2-4} 
                           & \begin{tabular}[c]{@{}l@{}}Tiny- \\ Imagenet\end{tabular} & 7.86 MB                                                    & 3.15 KB (0.04\%)                                                        \\ \hline
\end{tabular}

\caption{Results of detailed memory bandwidth overhead.}
\label{tab:overhead}
\end{table}


%


\subsection{Visualization}
Fig. \ref{fig:vis} shows the visualization of active or zero blocks in activation maps by overlapping them back to original images. As shown in the figure, many background blocks in input images do not contribute for classification. Model trained with Zebra can actually learn and skip these unimportant blocks efficiently. 

\section{Conclusion}
In this paper, we propose a zero block regularization method, Zebra, to help reduce the large memory bandwidth in modern CNN accelerators. Model trained with Zebra can dynamically use different amount of memory bandwidth according to different input images. Also, Zebra can be used with many different pruning method without additional efforts for even better results. The results show that model trained with Zebra can reduce 70\% of memory bandwidth for Resnet-18 on Tiny-Imagenet within 1\% accuracy drops and 2\% accuracy gain with the combination of Network Slimming.



\bibliographystyle{IEEEtran}

\bibliography{ieeeBSTcontrol,paper}

\end{document}